\documentclass[a4paper]{jpconf}
\usepackage{graphicx}
\usepackage{amsmath}
\usepackage{amssymb}

\newcommand{\fig}[1]{Fig.~\ref{fig:#1}}
\newcommand{\eq}[1]{(\ref{eq:#1})}

\newlength{\pictsize}
\begin{document}
\title{To the interpretation of the upturn observed by ATIC in heavy nuclei to iron ratios}

\author{%
A D Panov$^1$,
N V Sokolskaya$^1$, and
V I Zatsepin$^1$%
}

\address{Skobeltsyn Institute of Nuclear Physics, Moscow State University, Moscow, Russia}
\ead{panov@dec1.sinp.msu.ru}

\begin{abstract}
It is argued that the upturn observed in heavy nuclei to iron ratios as measured in the ATIC experiment could be understood within the model of closed galaxy with embedded local regions containing the sources of CR (Peters, Westergaard, 1977). The universal upturn near the energies of 200--300 GeV/n in the spectra of abundant primary nuclei is also predicted by this model, but it also predicts the source spectral index to be near $2.5$.
\end{abstract}

\section{Introduction}

The ratios of fluxes of nuclei from sulfur ($Z$=16) to chromium ($Z$=24) to the flux of iron  measured by the ATIC-2 experiment are presented in paper \cite{ATIC-2012-PANOV-ECRS1} at this conference. Hereinafter we call the range of charges \mbox{$Z$~=~16\,--\,24} as group $H^-$. The ratios are decreasing functions of energy from 5~GeV/n to approximately 80~GeV/n, as expected. However, an unexpected sharp upturns in the ratios are observed for energies above 100~GeV/n for all nuclei in group $H^-$ (Figs.~2--4 of \cite{ATIC-2012-PANOV-ECRS1}). No signs of systematic errors that may simulate the upturn in the ratio were found in the data, therefore the upturn was estimated as a real physical phenomenon, and it is meaningfull to discuss possible nature of the phenomenon. Due to a lack of the experimental data the discussion below should be considered as a very preliminary one.

\section{Simple leaky-box approximation}

It is suitable to start with simple leaky-box approximation to describe transport of particles. Let $N_1, N_2,\dots,N_k$ be secondary nuclei produced by spallation of iron. Then the ratio of summary flux of secondaries $I_{\Sigma S} = \sum_{i=1}^k I_{N_i}$ to the flux of iron may be written as
\begin{equation}
 \label{eq:SimpleLB}
 \frac{I_{\Sigma S}}{I_{\mathrm{Fe}}} = 
 \sum_{i=1}^k \dfrac{\varkappa_{N_i,\mathrm{Fe}}}{\varkappa^{N_i}_{\mathrm{esc}}(\varepsilon)+\varkappa_{N_i}},
\end{equation}
where $\varepsilon$ is the energy of particle per nucleon, $\varkappa^{N_i}_{\mathrm{esc}}=1/\lambda^{N_i}_{\mathrm{esc}}$ is the inverse diffusion escape length for leakage of the nucleus $N_i$ with energy $\varepsilon$ from the galaxy, $\varkappa_{N_i}=1/\lambda_{N_i}$ is the inverse nuclear spallation path length in the inter-stellar medium for the nucleus $N_i$, $\varkappa_{N_i,\mathrm{Fe}}=1/\lambda_{N_i,\mathrm{Fe}}$ is the inverse partial spallation path length of iron to produce the nucleus $N_i$. The escape length $\lambda_{\mathrm{esc}}$ is considered to be the universal function of rigidity for all nuclei and we use here the approximation from  \cite{NUCL-HEAO-1990-AA}: $\lambda_{\mathrm{esc}} = 34.1\,R^{-0.6}\, \mathrm{g}/\mathrm{cm}^2$.
\begin{figure}
\begin{center}
\includegraphics[width=\pictsize]{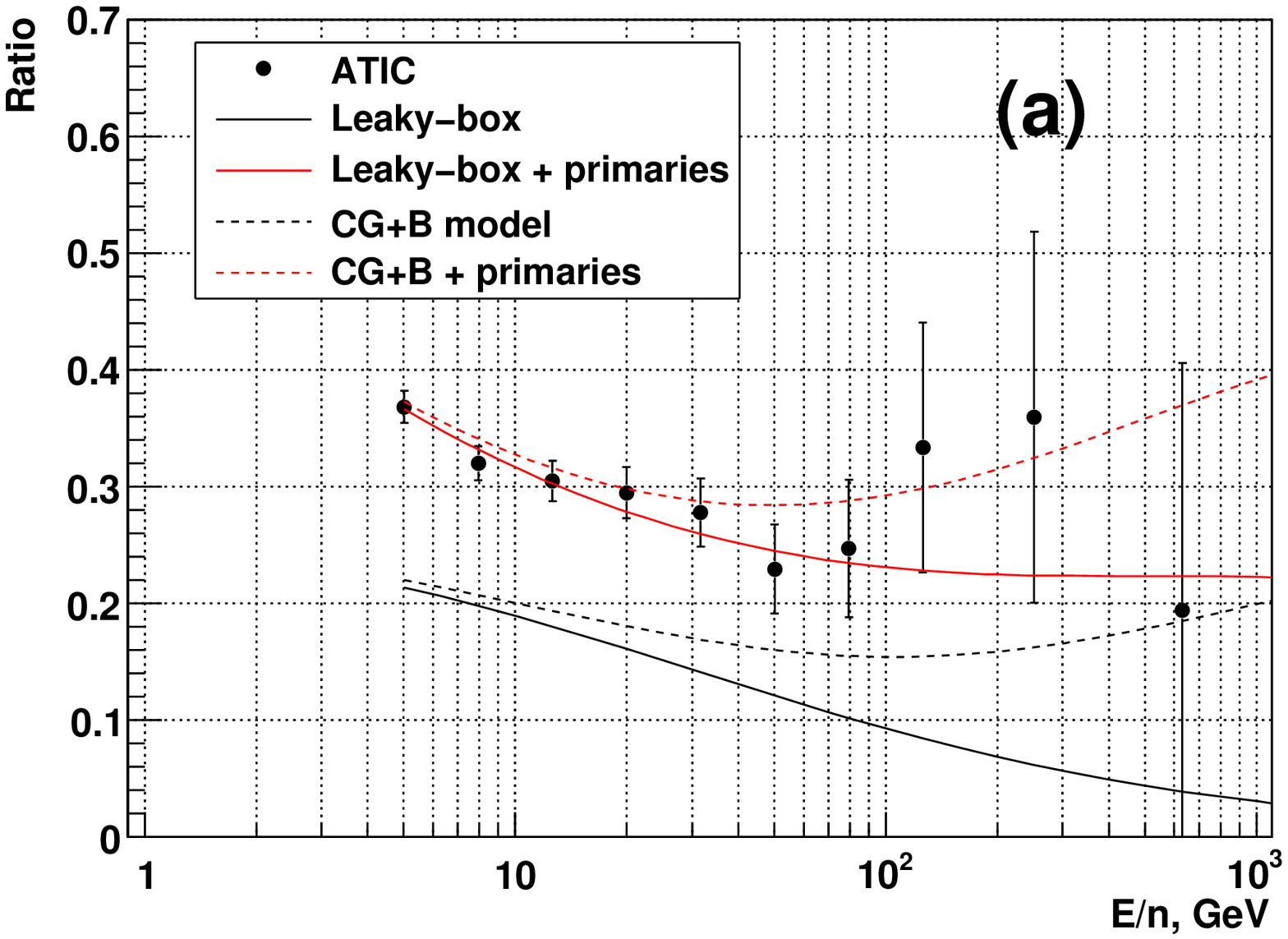}
\includegraphics[width=\pictsize]{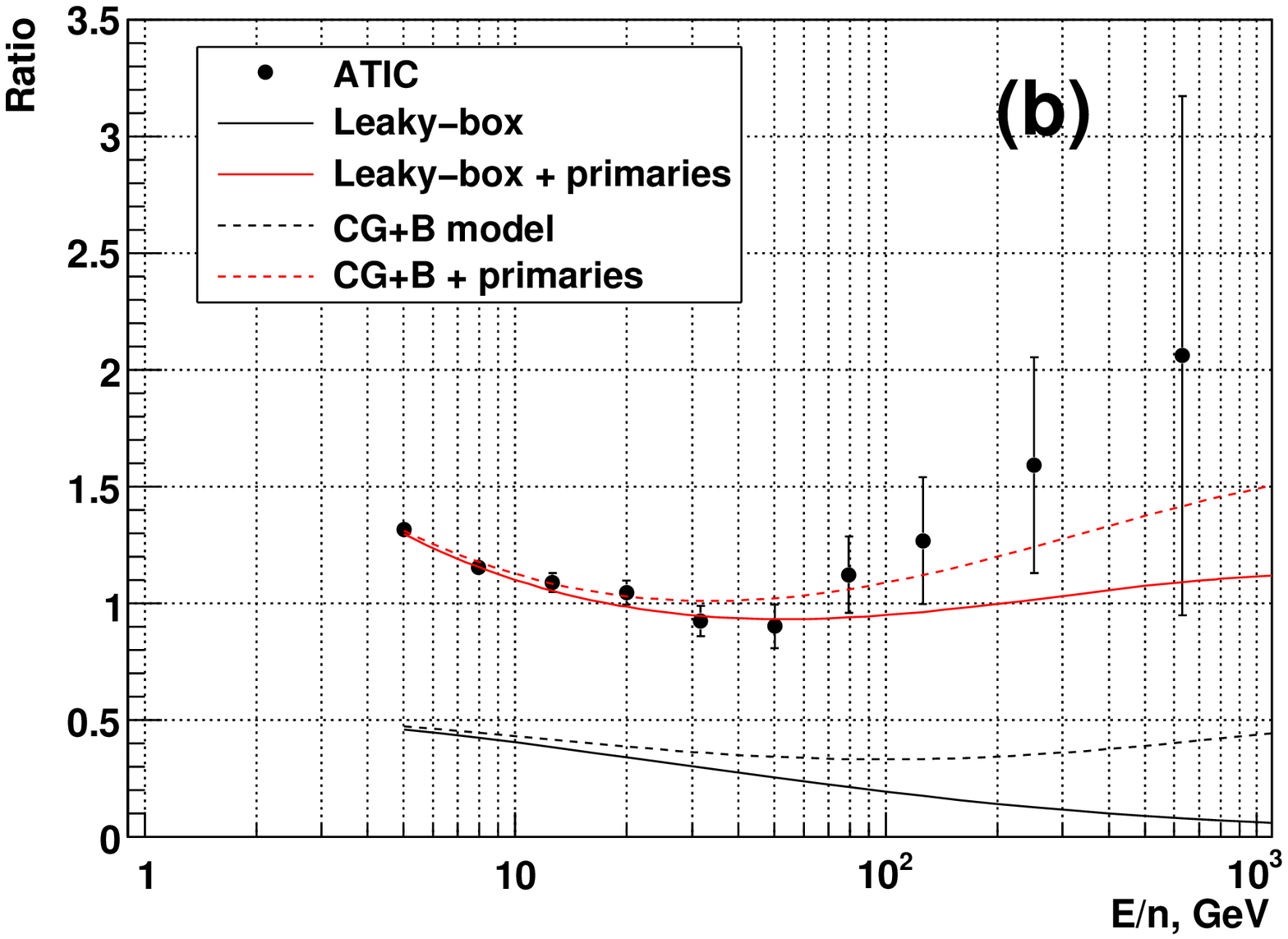}
\end{center}
\caption{\label{fig:Ratio-AllModels}Ratios of various nuclei groups to iron as measured by ATIC and a number of models (CG+B  means  Closed Galaxy with Bubbles, see the text). (a): (Ti+V+Cr)/Fe; (b): $H^-$/Fe.}
\end{figure}
We used the values of $\lambda_{N_i}$ compiled in the book of Ginzburg and Syrovatskii \cite{GINZBURG-1964}, and we calculated the partial path lengths $\lambda_{N_i,\mathrm{Fe}}$ with usage of the partial spallation cross sections compiled in \cite{SECT-FERRANDO-1988-PhysRev} and with supposition of 90\% protons and 10\% helium for interstellar medium.

We will study the flux (Ti+V+Cr) as a pattern of a group of nuclei with high contribution  of secondaries \cite{SHAPIRO1972-SpRes} and the flux of all $H^-$ group to obtain the highest statistics.  The ratio of fluxes (Ti+V+Cr)/Fe and $H^-$/Fe of the ATIC \cite{ATIC-2012-PANOV-ECRS1} experiment together with simple leaky-box model predictions (solid black lines) for secondary fluxes are shown in \fig{Ratio-AllModels}(a,b). The leaky-box model does not reproduce rising ratios at the energies $\varepsilon > 100$~GeV/n as could be expected. Leaky box model predicts also lower ratio in the range $\varepsilon < 40$~GeV/n than the ATIC data for the ratio (Ti+V+Cr)/Fe. It is a sign of some contribution of primary fluxes to the group Ti, V, Cr. This result is not new and had been obtained for Sc--Mn/Fe ratio in \cite{SECT-FERRANDO-1988-PhysRev} in 1988 (see FIG.~8 in \cite{SECT-FERRANDO-1988-PhysRev}). A prominent  contribution of the primary fluxes to S, Ar, Ca is expected.

\section{Addition of the primary fluxes to the model}

It is pointed out in \cite{ATIC-2012-PANOV-ECRS1} that the spectra of all abundant primary nuclei C, O, Ne, Mg, Si are very similar to each other and show an upturn in the ratio to the iron similar to the upturn in $H^-$/Fe but with lower amplitude. Suppose the same is true for primaries within the group $H^-$. Then one can fit the ratios (Ti+V+Cr)/Fe and $H^-$/Fe by the sum of the secondary flux predicted by the leaky-box model and a contribution of primaries. As a pattern for the shape of the primary spectrum we use the spectrum of oxygen (see \fig{PrimEvenClosedGalaxy}(b)). The ratio O/Fe was approximated by a cubic polynomial function of $\lg\varepsilon$. A fit of the ratio of sum of the leaky-box secondary spectrum and the primary spectrum to the iron flux for (Ti+V+Cr) and $H^-$ groups to the ATIC data are shown in \fig{Ratio-AllModels}(a,b) by red solid lines. There is a reasonable fit of the data at energies $\varepsilon < 100$~GeV/n, but there are no sufficient rising of the ratio at energies $\varepsilon \gtrsim 100$~GeV/n. 

\section{Model of closed galaxy with local bubbles}

Thus, the simple leaky-box model together with an addition of primary fluxes does not reproduce the data. Some additional ideas are needed to understand the upturn of the ratios. One possibility is the model of `closed galaxy' proposed in \cite{CRPROP-PETERS1977}. It was shown \cite{CRPROP-PETERS1977} that such a model may provide upturn in the sub-iron/iron ratio for secondary fluxes. Note that the model of closed galaxy was used in \cite{CREMULS-JACEE-1998-NuclPhysB} in the discussion of anomalously high ratio of Li+Be+B/C+N+O fluxes in the JACEE data.

It is supposed in the closed galaxy model that there are a number of compact regions in the galaxy each contains sources of CR and described by simple leaky-box model in the relation of diffusion leakage of particles from the region. Moreover, it is supposed all the CR sources are concentrated in such local regions. It was supposed in the original paper \cite{CRPROP-PETERS1977} that the compact regions were the galaxy arms, but they may be as well super-bubbles produced by supernova explosions. The last possibility looks reasonable if supernova explode mainly within the star associations where star formation process occurred shortly before, and massive short-living stars were created. The idea that super-bubbles may play important role in the forming of spectra of cosmic rays was widely discussed (see \cite{STREITMATTER2005-ICRC,EW2012A} and references herein). The exact nature of the local regions does not matter for the model but we will suppose for definiteness the super-bubbles and will call the model as `closed galaxy with bubbles' (CG+B). The second supposition of the model is that the whole galaxy is closed in the relation to the diffusion leakage. It is supposed in the model that the Sun is located in a Local Bubble, and the object of the model is to predict the CR fluxes in the Local Bubble.

The total flux of the CR in a bubble is comprised then of two parts \cite{CRPROP-PETERS1977}: 1) the local flux which may be described by usual leaky-box model applied to the bubble and 2) the global equilibrium galaxy flux (hereinafter we call it as a bulk flux) which also may be described by the model similar to the leaky-box model applied to the whole galaxy but with supposition of $\lambda_{\mathrm{esc}}(\varepsilon) \simeq \infty$. There is also one free parameter in such a model that represent the fraction of the bulk flux in the total flux which is unknown apriory and should be fitted to describe the data. 

The only sources of CR for the bulk within the CG+B model are surfaces of the bubbles. With usual supposition that the probability for a particle to leave a volume does not depend on already passed way within the volume one could obtain the equation for the modified bulk source:
\begin{equation}
 \label{eq:QBulk}
 Q_{\mathrm{bulk}}(\varepsilon) = 
 \frac{\varkappa_{\mathrm{esc}}(\varepsilon)}{\varkappa_{\mathrm{esc}}(\varepsilon) + \varkappa}\ Q(\varepsilon).
\end{equation}
Here $Q(\varepsilon)$ is the spectrum of some nucleus in the source within the bubble, $\varkappa_{\mathrm{esc}}(\varepsilon)$ is inverse length to escape for the nucleus from the bubble, and $\varkappa$ is inverse nuclear length for this nucleus. Using the equation \eq{QBulk} and usual formulae of leaky-box approximation applied to the bubble and to the galaxy one may obtain the ratio of total flux of a secondary nuclei $N_i$ to the flux of iron in the CG+B model as
\begin{equation}
 \label{eq:INkIFeRatio}
 \dfrac{I_{N_i}(\varepsilon)}{I_{\mathrm{Fe}}(\varepsilon)} = 
 \dfrac%
 {\dfrac{\varkappa_{N_i,\mathrm{Fe}}}{\varkappa^{N_i}_{\mathrm{esc}}(\varepsilon) + \varkappa_{N_i}} +
    K\ \dfrac{\varkappa_{N_i,\mathrm{Fe}}}{\varkappa_{N_i}}\
    \dfrac{\varkappa^{\mathrm{Fe}}_{\mathrm{esc}}(\varepsilon)}{\varkappa_{\mathrm{Fe}}}}%
 {1 + K\ \dfrac{\varkappa^{\mathrm{Fe}}_{\mathrm{esc}}(\varepsilon)}{\varkappa_{\mathrm{Fe}}}},
\end{equation}
where $K$ describes the fraction of the bulk flux in the total flux. For a flux of a group of nuclei one should sum by $i$ in the equation \eq{INkIFeRatio}.

\begin{figure}
\begin{center}
\includegraphics[width=\pictsize]{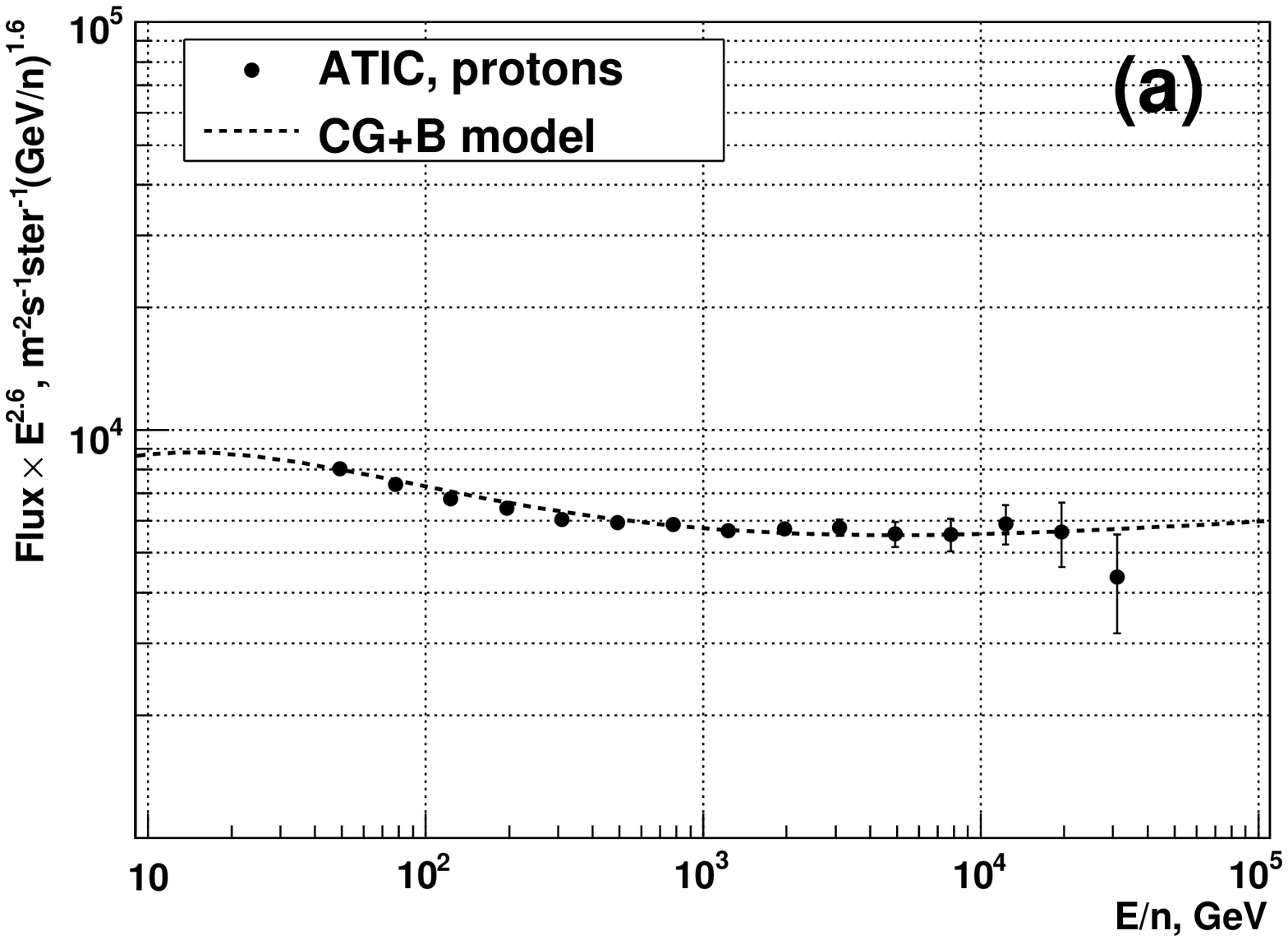}
\includegraphics[width=\pictsize]{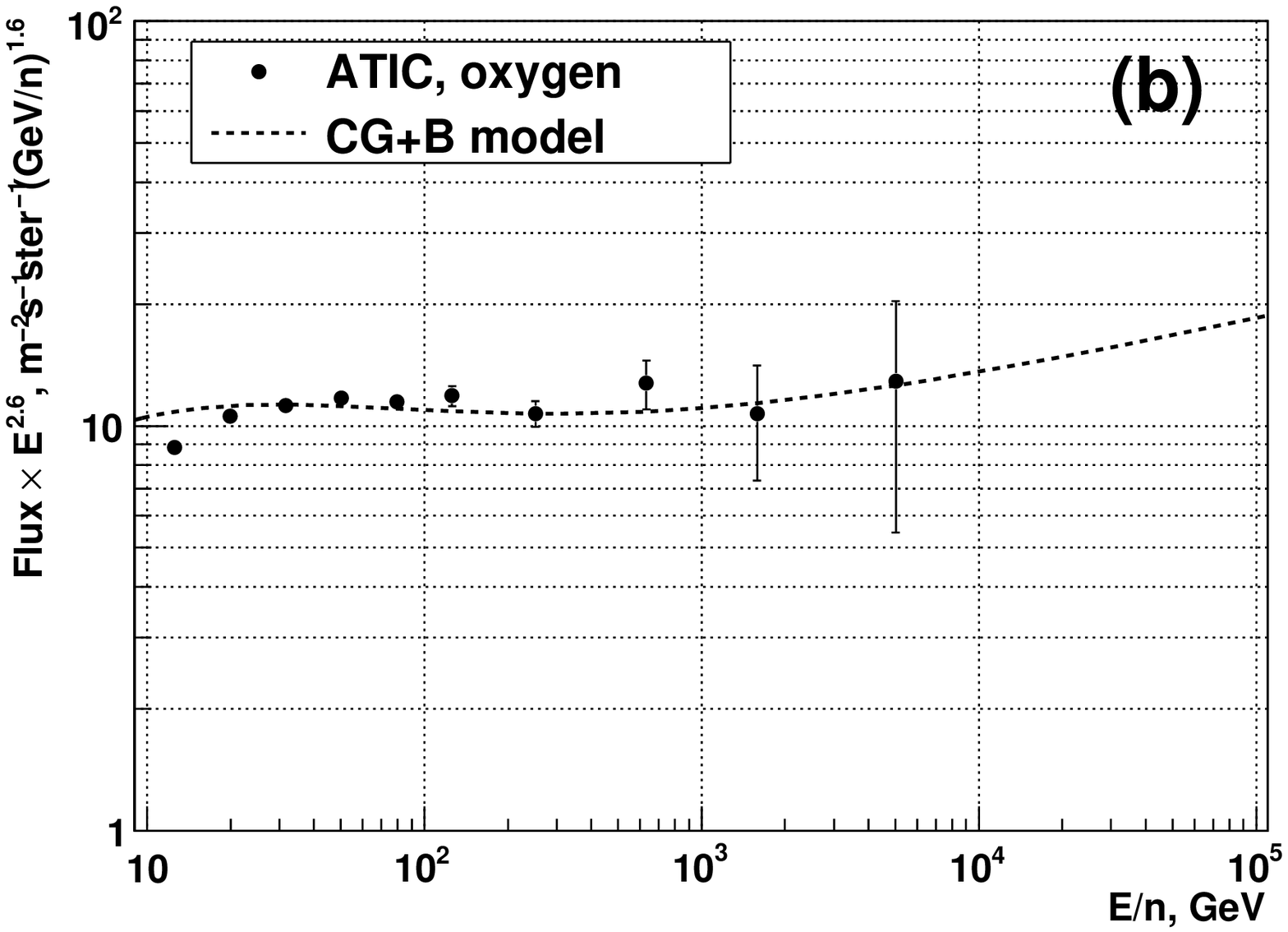}\\
\end{center}
\caption{Fit of abundant nuclei spectra measured by  ATIC-2  with the CG+B model: (a): protons, $\alpha_{\mathrm{source}}=2.55$; (b); oxygen, $\alpha_{\mathrm{source}}=2.45$.}
\label{fig:PrimEvenClosedGalaxy}
\end{figure}

The predictions of CG+B model for the fluxes of secondaries in the groups (Ti+V+Cr) and $H^-$ with $K=0.2$ are shown in \fig{Ratio-AllModels}(a,b) by dashed black lines. The complicated behavior of the model with its decreasing and increasing regions of ratios is a result of a concurrence of the local and bulk fluxes. Dashed red lines show CG+B model prediction together with the contribution  of primary fluxes exactly as it was done for the simple leaky-box models (solid red lines, \fig{Ratio-AllModels}). It is seen that the CG+B model together with the contribution of primary fluxes could describe the data in a reasonable approximation but the upturn of ratio in the experimental data may be more sharp than the predicted by the model one.

The spectra of primary abundant nuclei also could be understood within CG+B model. The fits of the proton and oxygen ATIC spectra by CG+B model are shown in \fig{PrimEvenClosedGalaxy}(a) and \fig{PrimEvenClosedGalaxy}(b) respectively. We have to suppose the source spectral indexes $\alpha(\mathrm{protons})=2.55$ and $\alpha(\mathrm{oxygen})=2.45$ to fit the data. These are very soft primary spectra and it may mean a problem for the CG+B model. However, a preference of the model is the prediction of the universal hardening of the spectra of abundant nuclei near 200--300~GeV/n similar to discovered by ATIC \cite{ATIC-2007-PANOV-IzvRan} and confirmed by CREAM \cite{CREAM2009-ApJ} without a hypothesis of additional special soft source spectrum of CR.

Note one important physical consequence of the CG+B model. Within the usual leaky-box model the diffusion coefficient may be estimated as \cite[p.~124]{GAISSER1990}: $D(\varepsilon) \sim \rho c H^2/\lambda_{\mathrm{esc}}(\varepsilon)$, where $H$ is some characteristic size of the system. In the case of the usual leaky-box model applied to the Galaxy, $H$ means the half-width of the galaxy halo (1--4~pc), but in the context of CG+B model $H$ means the half-size of the Local Bubble ($\sim100$~pc \cite{FRISCH1983,BERGHOEFER2002}). Since $\lambda_{\mathrm{esc}}$ is the same in both cases, and the expected half-size of the bubble is much less than the half-width of the halo and $\rho$ within the bubble expected to be much less than mean density in the Galaxy \cite{FRISCH1983,BERGHOEFER2002}, then CG+B predicts diffusion coefficient much less (2--3 orders of magnitude or even more) than usually adopted value. This estimate supposes free escape of particles from the margin of the bubble. The alternative explanation of the confinement of cosmic rays in the Local Bubble is reflection of particles by the walls of the bubble \cite{EW2012A} or it may be some combination of low-value diffusion coefficient and reflection by the walls.

%Let us emphasize again that we have taken only a first attempt to understand the observed upturn in heavy nuclei to iron ratios. 
%A criticism and discussion of the proposed above simple models may promote better understanding of the phenomenon.

The work was supported by RFBR grant number 11-02-00275.

\section*{References}
% \bibliography{SubIronBibl}

\providecommand{\newblock}{}

\end{document}